\documentclass[twocolumn, amsmath, superscriptaddress, amsfonts,prb]{revtex4}
\pdfoutput=1
\usepackage{graphicx}
\begin{document}
\title{The stochastic growth of metal whiskers}
\author{Biswas Subedi}\email{biwas.subedi@rockets.utoledo.edu}\affiliation{Department of Physics and Astronomy, University of Toledo, Toledo,OH 43606, USA}
\author{Dipesh Niraula}\email{dipesh.niraula@rockets.utoledo.edu}\affiliation{Department of Physics and Astronomy, University of Toledo, Toledo,OH 43606, USA}
\author{Victor G. Karpov}\email{victor.karpov@utoledo.edu}\affiliation{Department of Physics and Astronomy, University of Toledo, Toledo,OH 43606, USA}

\begin{abstract}
The phenomenon of spontaneously growing metal whiskers (MW) raises significant reliability concerns due to its related arcing and shorting in electric equipment. The growth kinetics of MW remains poorly predictable. Here we present a theory describing the earlier observed intermittent growth of MW as caused by local energy barriers related to variations in the random electric fields generated by surface imperfections. We find the probabilistic distribution of MW stopping times, during which MW growth halts, which is important for reliability projections.

\end{abstract}

\maketitle

\begin{figure}[t!]
\includegraphics[width=0.47\textwidth]{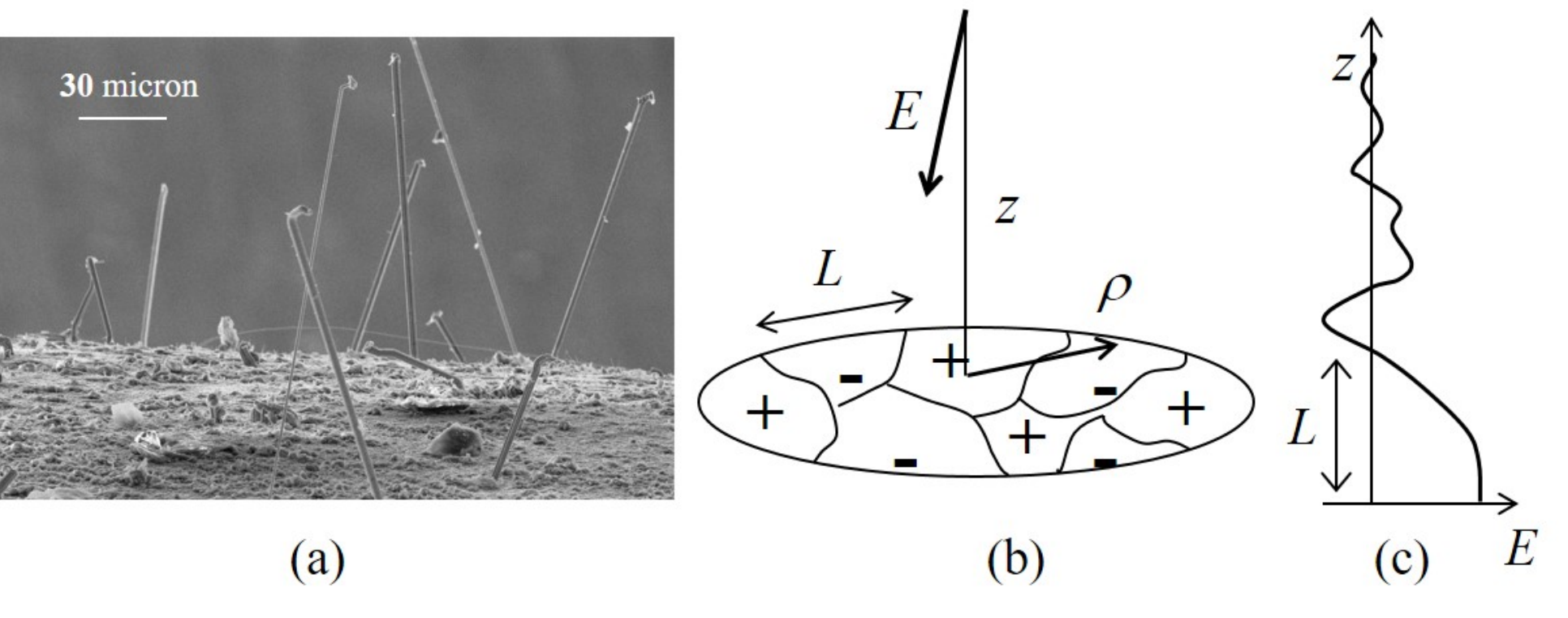}
\caption{(a) Scanning electron microscope (SEM) pictures of
zinc whiskers. Courtesy of the NASA Electronic
Parts and Packaging (NEPP) Program. (b) A sketch of charge patches on a metal; surface and their induced random electric field. (c) A sketch of the coordinate dependence of the random electric field vs. the distance from a metal surface.}
\label{Fig:field}\end{figure}

{\it Introduction} -- Metal whiskers [MW; illustrated in Fig. \ref{Fig:field} (a)] are hair-like protrusions growing from the surfaces of many metals, such as Sn, Zn, Cd, and Ag. MW caused shorting in electronic packages raises significant reliability concerns and losses in different technologies ranging from aerospace and military to auto industry and medical devices. \cite{NASA1,barnes,galyon2003,brusse2002,panashchenko2009} After about 70-year of observations, the understanding of MW growth remains insufficient.
\begin{figure}[t!]
\includegraphics[width=0.40\textwidth]{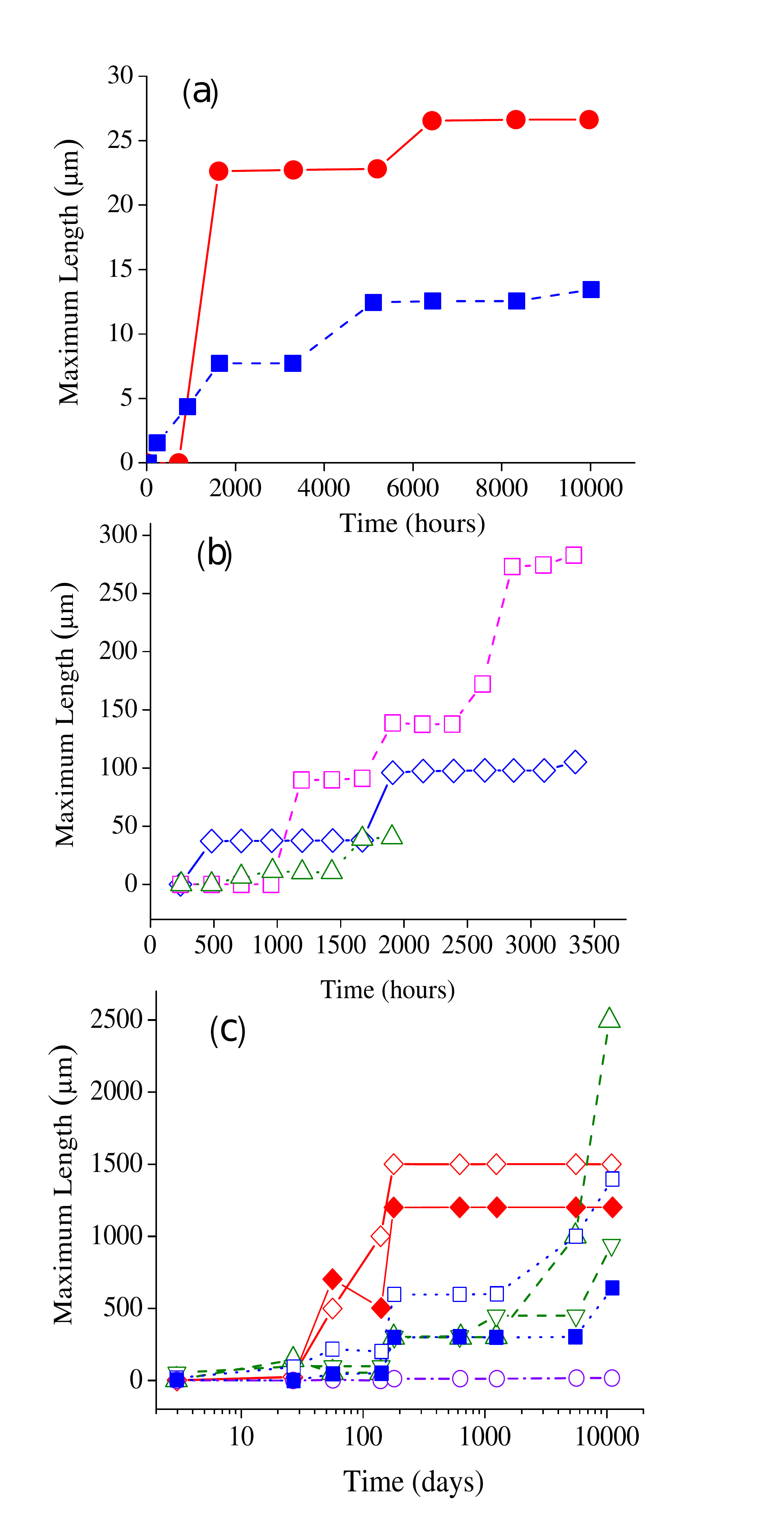}
\caption{Compilation of published data on the intermittent MW growth for different capping, environmental and stress conditions (unspecified here): (a) different capping, \cite{kim2008} (b) different environments, \cite{meschter2015} (c) various substrates and stresses. \cite{ashworth2016} Several domains showing a decrease of MW length in graph (c) are artifacts related to the problems with identification of the maximum length whisker.\cite{dunn2017}  }
\label{Fig:data}\end{figure}

Mechanical stresses, \cite{sarobol2013,pei2013,pei2014, chason2014}  local recrystallization regions, \cite{vianco2015, qiang2014} intermetallic compounds, \cite{tu1994,so1996} and stress gradients\cite{stein2014,yang2008,sobiech2008,sobiech2009} have been considered as MW driving forces.   A recent electrostatic concept attributes MW growth to random electric fields generated by surfaces of imperfect metals. \cite{shvydka2016,karpov2014,karpov2015,niraula2015,vasko2015,vasko2015a,borra2016,niraula2016}

The predictability of MW effects is hindered by their stochastic nature: MW lengths ($h$) and diameters ($d$) are mutually uncorrelated and broadly distributed obeying log-normal statistics; the local concentration of MW varies exponentially between different regions. Many other observations summarized by G. Davy \cite{davy2014} indicate that stochasticity as well. In particular, MW ``...growth rate is often not constant. A whisker may stop growing for a while, then start growing again." \cite{davy2014,kostic2014} MW growth randomly interrupted for years \cite{ashworth2016} and days \cite{kim2008,meschter2015} was observed; Fig. \ref{Fig:data} presents a compilation.

Here, we present a theory describing the stochastically intermittent longitudinal growth of MW.  We establish the statistics of MW barriers ($V$) and stopping times,
\begin{equation}\label{eq:tau}\tau=\tau _0\exp\left(V/kT\right),\end{equation} during which the longitudinal MW growth ceases; here $\tau _0={\rm const}$ and $kT$ is the thermal energy. The practical side of our work is that the time dependent probabilities of MW lengths determine the likelihood of their related reliability failures.

{\it Model} --Our consideration is based on the electrostatic theory, \cite{karpov2014,niraula2015,shvydka2016} according to which the major factor behind MW growth is the random electric field generated by charged surface imperfections, such as grain boundaries, contaminations, chemical or structural nonuniformities. The random distribution of charges is presented by the uncorrelated charge patches of a certain dimension $L$ illustrated in Fig. \ref{Fig:field}.

\begin{figure}[t]
\includegraphics[width=0.5\textwidth]{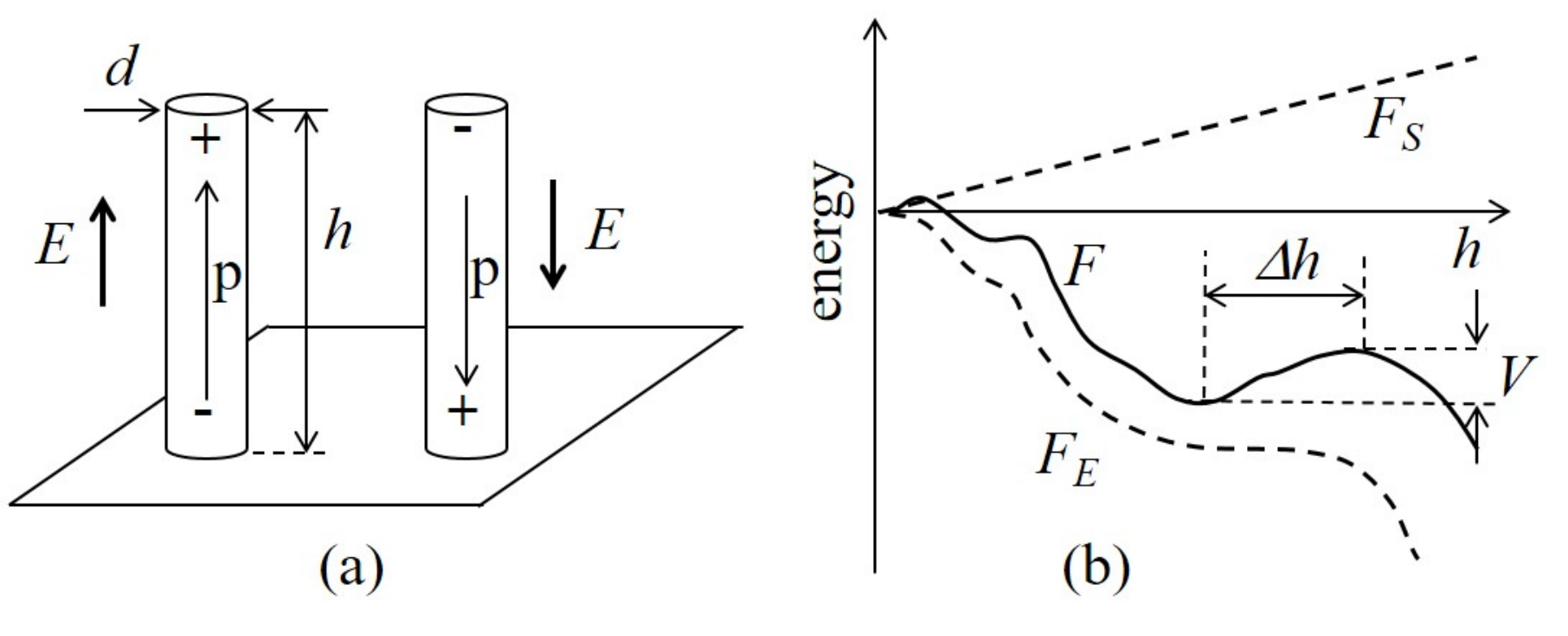}
\caption{(a) A model of metal whiskers showing two metal cylinders in the areas of opposing electric fields $E$ inducing the corresponding electric dipoles $p$, such that the electrostatic energy gain $-pE$ does not depend on the sense of the electric field. (b) A sketch of the free energy vs. MW length showing a stopping barrier of height $V$.}
\label{Fig:energy} \end{figure}

As sketched in Fig. \ref{Fig:energy} (a), the field induced electric dipole will decrease MW energy by $-pE=-\beta E^2$ regardless of the field orientation. Because the polarizability of a needle shaped metal particle \cite{landau1984} $\beta\sim h^3$ can be rather high while its surface is small, MW nucleation and growth become possible. The free energy of MW can be written as,
\begin{equation}\label{eq:freeen1}
F=-\frac{1}{2\Lambda}\int_{0}^{h}\xi^2dh+\pi\sigma hd \quad {\rm with}\quad \xi\equiv\int_{0}^{h}Edz
\end{equation}
where the first and second terms represent respectively the electrostatic ($F_E$) and surface ($F_S$) contributions, $\sigma$ is the surface tension, $d$ is MW diameter, $h$ is its length, $E$ is the normal (along $z$-coordinate parallel to the whisker axis) component of the random electric field, and $\Lambda\ = \ln(4h/d)-1\gg 1$. Because the electric field $E(z)$ is random, so are the functionals $\xi$, $F_E$, and $F$. In particular, randomly located regions of low field strength give rise to the surface energy related barriers that can either temporarily or permanently inhibit whisker growth. The effect of such barriers was shown to predict the log-normal distribution of MW lengths. \cite{karpov2014,niraula2015} However, related to stopping times, the probabilistic distribution of that barrier heights remained unknown.

To analytically describe a stopping barrier, we use the Taylor expansion in the proximity of an arbitrary MW length $h_0$
\begin{equation}\label{eq:taylor}
F(h)={\rm const}+ ax+bx^2+cx^3,\quad x\equiv h-h_0>0,\end{equation}
which yields a barrier height and width,
\begin{equation}\label{eq:VB}
 V=\frac{4(b^2-3ac)^{3/2}}{27c^2}, \quad {\rm and}\quad \Delta h=\left(\frac{V}{4|c|}\right)^{1/3}.
\end{equation}

{\it Statistical Analysis} -- The probabilistic properties of such stopping barriers are determined by the statistics of  random coefficients, $a$, $b$, and $c$, which can be expressed from Eq. (\ref{eq:freeen1}),
\begin{equation}\label{eq:a}
a= -\frac{\xi^2}{2\Lambda}+\sigma \pi d,\  b=-\frac{f\xi}{2\Lambda},\ c=-\frac{1}{12\Lambda}\left(\xi\frac{\partial f}{\partial h}+f^2\right)
\end{equation}
where $f\equiv \partial\xi/\partial h$, and we have neglected relatively small derivatives of the logarithmically weak function $\Lambda$.

Using Coulomb's law the random quantities $\xi$ and $f$ can be expressed through the random surface charge density $en({\boldsymbol\rho})$ where $e$ is the elemental charge,
\begin{equation}\label{eq:xi}
\xi = \int_{0}^{h}dz\int \frac{en(\boldsymbol\rho)zd^{2} \rho}{(z^{2}+\rho^{2})^{3/2}}, \ f = \int \frac{en(\boldsymbol\rho)hd^2\rho}{(h^2+\rho^2)^{3/2}}
\end{equation}
where $\boldsymbol\rho$ is the 2D radius vector.
Being integrals of large numbers of random contributions, $\xi$ and $f$ satisfy the conditions of central limit theorem and obey Gaussian distributions.

The latter distributions' moments are readily effected for the case of delta-correlated surface charges,
\begin{equation}\label{eq:correlator}
e^2\langle n(\boldsymbol\rho)n (\boldsymbol\rho ')\rangle=C\delta (\boldsymbol\rho -\boldsymbol\rho ') \quad {\rm when}\quad \rho\gg L.
\end{equation}
Here the angular brackets represent averaging, $\delta$ stands for the delta-function, and
\begin{equation}\label{eq:coefC}C\sim (\overline{n}eL)^2\end{equation} is a constant where $\overline{n}$ is the characteristic rms fluctuation of surface charge density. The delta-function representation remains adequate when $\rho\gg L$ where $L\sim 0.1-1$ $\mu$m is the characteristic linear dimension of a charge patch. Using Eq. (\ref{eq:correlator}), the definitions in Eq. (\ref{eq:xi}), and assuming long enough whiskers, $h\gg L$, yields
\begin{equation}\label{eq:disp2}
\langle\xi ^2\rangle =2\pi C\ln\frac{h}{4L},\quad  \langle f^2\rangle=\frac{\pi C}{2h^2}, \quad  \langle \xi \frac{\partial f}{\partial h}\rangle=-\frac{3\pi C}{2h^2}.
\end{equation}

We are now in a position to describe the probabilistic distribution of stopping barriers. Substituting Eq. (\ref{eq:a}) into Eq. (\ref{eq:VB}) yields,
\begin{equation}\label{eq:VB1}
V=\frac{4}{3\sqrt{2}\Lambda}\frac{(\xi \sqrt{\zeta}-\xi _c)^{3/2}(\xi \sqrt{\zeta}+\xi _c)^{3/2}}{(-\xi\partial f/\partial h-f^2)^{1/2}}
\end{equation}
where we have introduced
\begin{equation}\label{eq:zeta}
\zeta \equiv\frac{\xi\partial f/\partial h-f^2}{\xi\partial f/\partial h+f^2}\quad {\rm and}\quad\xi_c\equiv \sqrt{2\pi\Lambda\sigma d}.
\end{equation}

The random quantities $f^2$ and $\xi\partial f/\partial h$ in Eq. (\ref{eq:VB1}) and (\ref{eq:zeta}) are approximated by their averages from Eq. (\ref{eq:disp2}): using the same technique as in Appendix B of Ref. \onlinecite{karpov2014}, it is seen that their rms fluctuations are proportional to the small parameter $L/h\ll 1$. In that approximation, one gets, $\zeta =2$, and
\begin{equation}\label{eq:VB2}
V=V_{0}(\tilde{\xi} -\tilde{\xi _c})^{3/2}({\tilde\xi} +\tilde{\xi _c})^{3/2}
\end{equation}
with
\begin{equation}\label{eq:norm}
V_{0}=\frac{16\sqrt{2}\pi hC[\ln(h/4L)]^{3/2}}{3\Lambda},\quad \tilde{\xi}=\frac{\xi}{\sqrt{\langle\xi ^2\rangle}}
\end{equation}
and
\begin{equation}\label{eq:norm1}
\tilde{\xi _c}=\sqrt{\frac{\alpha}{\ln(h/4L)}}, \quad \alpha\equiv\frac{\Lambda\sigma d}{2C}.
\end{equation}

Using Eqs. (\ref{eq:norm}) and (\ref{eq:coefC}) the characteristic barrier $V_{0}$ is estimated as,
\begin{equation}\label{eq:VB0}\frac{V_{0}}{kT}\sim \frac{e^2}{LkT}\frac{h}{L}(\overline{n}L^2)^2\sim 10^7-10^9\end{equation}
where we have used \cite{karpov2014} $L\sim 0.1-1$ $\mu$m, $\overline{n}\sim 10^{12}$ cm$^{-2}$, and $h/L\sim 100$. In particular, short enough stopping times $\tau$ possible for observations correspond to $V\ll V_{0}$, thus $\tilde{\xi}$ very close to $\pm\tilde{\xi _c}$.
The condition of low barriers, $V\ll V_{0}$ defines simultaneously the range of applicability of approximation in Eq. (\ref{eq:taylor}). Indeed, replacing $c$ in the expression for $\Delta h$ in Eq. (\ref{eq:VB}) with its average by virtue of Eq. (\ref{eq:a}) yields,
\begin{equation}\label{eq:dh1}\Delta h=h(6\Lambda /\pi)^{1/3}(V/Ch)^{1/3}.\end{equation}

Expressing $\tilde{\xi}$ from Eq. (\ref{eq:VB2}),
the Gaussian distribution $\exp(-\tilde{\xi}^2/2)$ yields the normalized barrier distribution,
\begin{equation}\label{eq:rhoVB}
\rho (V)=\frac{2}{3V^{1/3}V_0^{2/3}} \exp\left[-\left(\frac{V}{V_{0}}\right)^{2/3}\right]\end{equation}
at distance $h$ from the metal surface [recall that $V_0$ depends on $h$ as specified in Eq. (\ref{eq:norm})].
Because for all practical cases one has to assume $V\ll V_0$, the distribution in Eq. (\ref{eq:rhoVB}) is not very different from uniform, $\rho (V)\approx {\rm const}\sim 1/V_0$, which is typical of many models of disordered systems [note that $V_0$ is length dependent in Eq. (\ref{eq:norm})].

The corresponding distribution of stopping times is,
\begin{equation}\label{eq:rhotau}g(\tau )=(kT/\tau )\rho [kT\ln(\tau /\tau _0)].\end{equation}
Neglecting again the exponent, the latter distribution can be closely approximated by the generic form $g(\tau )\propto 1/\tau$ known for many disordered systems.

There is a strong correlation between the random quantities corresponding to the same length scale,
\begin{eqnarray}\label{eq:corrco}
r&\equiv &\frac{\langle\xi (h)\xi (h')\rangle }{\sqrt{\langle\xi ^2(h)\rangle\langle\xi ^2(h')\rangle}} =\frac{\ln[hh'/2(h+h')L]}{\sqrt{\ln(h/4L)\ln(h'/4L)}} \nonumber \\ &\approx & 1-\frac{1}{8\ln(h/4L)}\left(\frac{ h'-h}{h}\right)^2,
\end{eqnarray}
and similar for other correlation coefficients, where the latter equality represents the case of small $|h-h'|\ll h$ and we have used Eqs. (\ref{eq:xi}) and (\ref{eq:correlator}).

{\it Conclusions} --The following conclusions can be made based on the results in Eqs. (\ref{eq:norm}), (\ref{eq:rhoVB}), (\ref{eq:rhotau}) and (\ref{eq:corrco}). \\
1) The electrostatic theory naturally predicts the stochastic intermittent kinetics of MW growth.\\
2) For all practical purposes, the stopping barriers can be described in the cubic approximation.\\
3) Their probabilistic distribution is close to uniform. \\
4) The stopping time distribution is close to $g(\tau )\propto 1/\tau$ characteristic of many processes in disordered systems. It predicts that short time interruptions are more likely and can be overlooked unless intentionally tracked. \\
5) The barriers typically extend over the entire length ranges at which they appear; the next barrier for the same MW is expected when its length changes by an order of magnitude. \\
6) The characteristic barrier height $V_0\propto h$ increases with MW length [Eq. (\ref{eq:norm})], and the characteristic stopping time increases with $h$ exponentially.

Our theory relates the intermittent growth of MW to the physics of disordered systems where random barriers inhibit a system development. It neglects the details of the underlying material structure and diffusion emphasizing instead the general disorder effects. An alternative interpretation referring to material specific aspects can suggest that ``...As the tin plating anneals and grain growth occurs there will be interruptions in the mechanism of whisker growth, particularly when grain annihilation or abnormal grain growth takes place during recrystallization. This could account for `stop-and-go' growth."\cite{dunn2017}

Collecting sufficient data on the intermittency statistics, such as the stopping time $\tau$ distribution and correlations between $\tau$ and MW length $h$ could help to experimentally verify the above theory.

We would like to summarize our consideration by saying that the occasionally observed intermittent growth is explained by the electrostatic theory accounting for barriers in MW free energies. In our stochastic picture of MW development, short time growth interruptions happen more often, especially at the earlier stages, and can be overlooked. On the other hand, longer MW can show exponentially longer interruptions that can be misinterpreted as the end of MW growth.

Industry standards for MW propensity limited time tests, such as suggested by Joint Electron Devices Engineering Council, \cite{industry,comments} need to be approached statistically, predicting {\it with a certain probability} the long-term growth of MW, for example, the probability of a certain MW growth during the desired time interval.  Industrial protocols for such statistical predictions can be attempted based on a theory such as the above quantifying possible losses in the spirit of actuarial analyses.

The authors acknowledge useful discussions with D. Shvydka, A. D. Kostic, G. Davy, J. Brusse, B. Dunn, S. Meschter, and P. Snugovsky.

\end{document}